\documentclass[12pt,a4]{article}
\usepackage{amsmath}
\usepackage{bm}
\usepackage{amsfonts}
\usepackage{amssymb}
\usepackage{graphics}
\usepackage[normal]{caption2}
\usepackage{subfigure}
\usepackage{rotating}
\usepackage{citesort}
\def\be{\begin{equation}}
\def\ee{\end{equation}}
\def\ba{\begin{array}}
\def\ea{\end{array}}
\def\bea{\begin{eqnarray}}
\def\eea{\end{eqnarray}}

\begin{document}
\baselineskip 20pt \setlength\tabcolsep{2.5mm}
\renewcommand\arraystretch{1.5}
\setlength{\abovecaptionskip}{0.1cm}
\setlength{\belowcaptionskip}{0.5cm}
\begin{center} {\large\bf On the sensitivity of transverse flow to the isospin-dependent cross-section and symmetry energy}\\
\vspace*{0.4cm}
{\bf Sakshi Gautam and Rajeev K. Puri}\footnote{Email:~rkpuri@pu.ac.in}\\
{\it  Department of Physics, Panjab University, Chandigarh -160
014, India.\\}
\end{center}
We study the transverse flow for systems having various
 N/Z ratios. We find the transverse flow is sensitive to N/Z
 ratio and, in fact, increases with N/Z of the system. The
 relative contribution of symmetry energy and isospin dependence
 of nucleon-nucleon cross section is also investigated. We find the greater sensitivity of symmetry energy in the
 lighter
 systems compared to heavier ones.


\newpage
\baselineskip 20pt

Rapid advances in the technologies to accelerate radioactive ion
beams (RIBs) has opened up several new frontiers in nuclear
science. In particular, the availability of RIB facilities at
Cooler Storage Ring (CSR) (China) \cite{rib1}, the GSI Facility
for Antiproton and Ion beam Research (FAIR) \cite{rib2}, RIB
facility at Rikagaku Kenyusho (RIKEN) in Japan \cite{rib3}, GANIL
in France \cite{rib4}, and the upcoming facility for RIB at
Michigan State University \cite{rib5} provide
  great opportunity to explore the equation of state and properties of
 dense neutron-rich matter. This knowledge is important not only
 for the understanding of the structure of radioactive nuclei, it may also address
 several critical issues in astrophysics.
 \par
 The intensive studies by nuclear physics community led to the
 determination of the equation of state (EOS) of symmetric nuclear
 matter. In particular, the incompressibility of symmetric nuclear
 matter at densities greater than normal nuclear matter density
 has been constrained by the measurements of collective flows \cite{daniel}  in
 nucleus-nucleus collisions.
 \par
 The collective transverse in-plane flow has been used extensively
 over the past three decades to study the properties of hot and
 dense nuclear matter, i.e., nuclear EOS and in-medium
 nucleon-nucleon cross section. The study of isospin effects in
 collective transverse in-plane flow also helps to obtain
 information about isospin-dependent mean field. The isospin
 effects in collective transverse flow were first predicted by Pak
\emph{ et al}. \cite{pak}. These isospin effects in collective
flow has been
 explained  as the competition among various reaction mechanisms
 such as nucleon-nucleon collisions, symmetry energy, surface
 properties of the colliding nuclei and Coulomb force \cite{li}. In Ref. \cite{gaum10,gaum210}
one of us and collaborators predicted the isospin effects in flow
and its
 disappearance for the isobaric pairs. There we found the
 dominance of Coulomb repulsion over the symmetry energy in isospin
 effects. In another study of isotopic pairs, two of us and collaborators \cite{gaum310} have
 studied the effect of symmetry energy on the collective
 transverse flow of neutron-rich system of $^{60}$Ca+$^{60}$Ca.
 The study revealed the sensitivity of flow to the symmetry energy
 in the Fermi energy region and shows insensitivity at higher
 incident energies of 400 and 800 MeV/nucleon. In Ref. \cite{gaum410}
Gautam and Sood studied the N/Z dependence of energy of
disappearance of
 flow (energy at which transverse in-plane flow vanishes)
 throughout the mass range. The study pointed towards the role of
 symmetry energy in the N/Z dependence of energy of vanishing
 flow.  Motivated by all the previous studies, in the present
 paper, we aim to see the relative contribution of symmetry energy
 and isospin dependence of nucleon-nucleon cross section on the
 collective transverse in-plane flow. The present study is carried
 out within the framework of isospin-dependent quantum molecular
 dynamics (IQMD) model \cite{hart98}.

\begin{figure}[!t] \centering
 \vskip -1cm
\includegraphics[width=14cm]{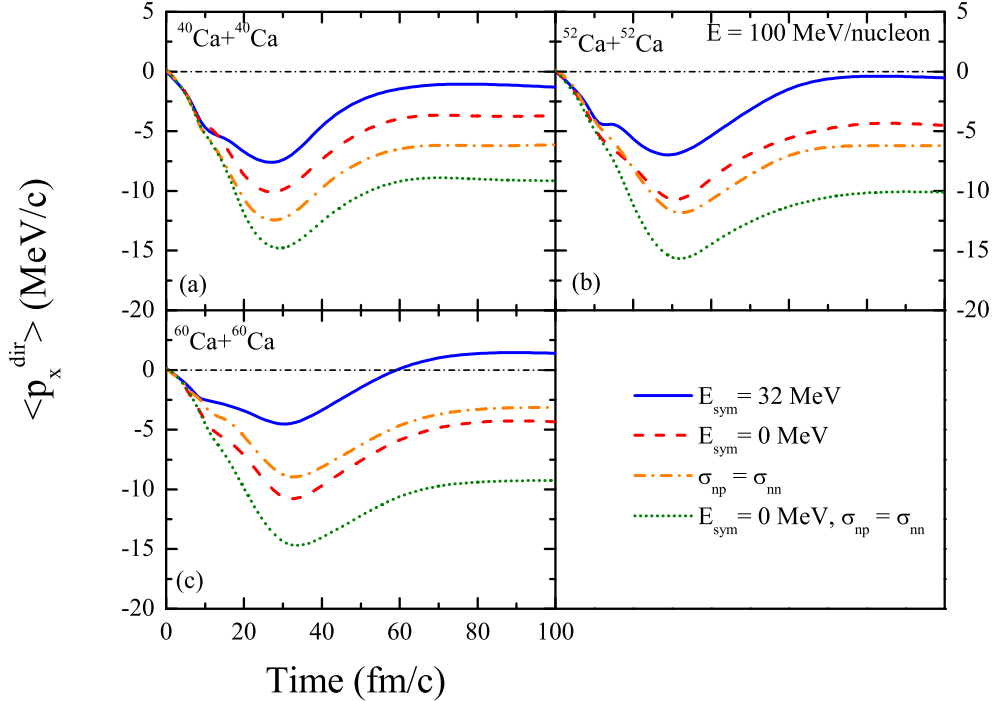}
\caption{(Color online) The time evolution of
$<p_{x}^{\textrm{dir}}>$ for the reactions of Ca+Ca having N/Z =
1.0, 1.6 and 2.0 at 100 MeV/nucleon. Lines are explained in the
text.}\label{fig1}
\end{figure}
\par
We simulate the reactions of Ca+Ca and Xe+Xe series having N/Z =
1.0, 1.6 and 2.0. In particular, we simulate the reactions of
$^{40}$Ca+$^{40}$Ca, $^{52}$Ca+$^{52}$Ca, $^{60}$Ca+$^{60}$Ca and
$^{110}$Xe+$^{110}$Xe, $^{140}$Xe+$^{140}$Xe and
$^{162}$Xe+$^{162}$Xe at impact parameter of
b/b$_{\textrm{max}}$=0.2-0.4. The incident energy is taken to be
100 MeV/nucleon. We use a soft equation of state along with the
standard isospin- and energy-dependent cross section reduced by
  20$\%$, i.e. $\sigma$ = 0.8 $\sigma_{nn}^{free}$.
The reactions are followed till the transverse in-plane flow
saturates. It is worth mentioning here that the saturation time
varies with the mass of the system. It has been shown in Ref.
\cite{sood1} that the transverse in-plane flow in lighter
colliding nuclei saturates earlier compared to heavy colliding
nuclei. Saturation time is about 100 (150 fm/c) in lighter (heavy)
colliding nuclei in the present energy domain. We use the quantity
"\textit{directed transverse momentum $\langle
p_{x}^{dir}\rangle$}" to define the nuclear transverse in-plane
flow, which is defined as \cite{sood1}
\begin {equation}
\langle{p_{x}^{dir}}\rangle = \frac{1} {A}\sum_{i=1}^{A}{sign\{
{y(i)}\} p_{x}(i)},
\end {equation}
where $y(i)$ and $p_{x}$(i) are, respectively, the rapidity
(calculated in the center of mass system) and the momentum of the
$i^{th}$ particle. The rapidity is defined as
\begin {equation}
Y(i)= \frac{1}{2}\ln\frac{{\vec{E}}(i)+{\vec{p}}_{z}(i)}
{{\vec{E}}(i)-{\vec{p}}_{z}(i)},
\end {equation}

where $\vec{E}(i)$ and $\vec{p_{z}}(i)$ are, respectively, the
energy and longitudinal momentum of the $i^{th}$ particle. In this
definition, all the rapidity bins are taken into account.

\begin{figure}[!t] \centering
 \vskip 1cm
\includegraphics[angle=0,width=14cm]{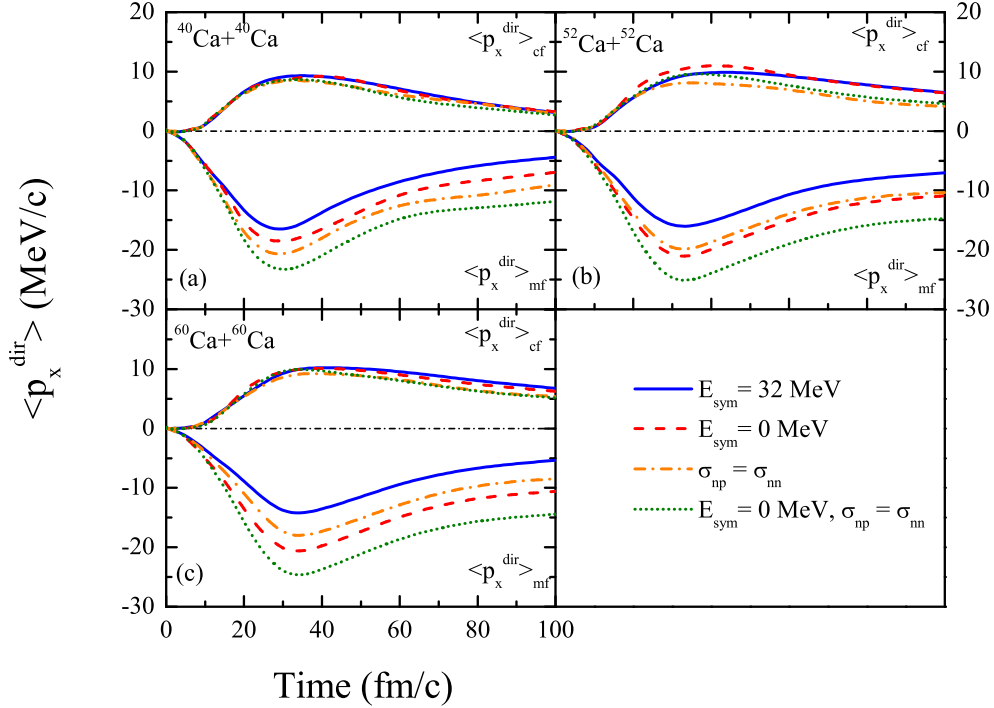}
 \vskip -0cm \caption{ (Color online) The decomposition of
$<p_{x}^{\textrm{dir}}>$ into $<p_{x}^{\textrm{dir}}>_{mf}$ and
$<p_{x}^{\textrm{dir}}>_{cf}$ for the reactions of Ca+Ca having
N/Z = 1.0, 1.6 and 2.0 at 100 MeV/nucleon. Lines have same meaning
as in Fig. 1.} \label{fig2}
\end{figure}

\par
In fig. 1, we display the time evolution of
$<p_{x}^{\textrm{dir}}>$ for the reactions of $^{40}$Ca+$^{40}$Ca,
$^{52}$Ca+$^{52}$Ca and $^{60}$Ca+$^{60}$Ca. We see that at the
 start of the reaction, $<p_{x}^{\textrm{dir}}>$ is
negative (due to the dominance of mean-field), reaches a minimum
and then increases and saturates at around 80 fm/c. The values of
$<p_{x}^{\textrm{dir}}>$ is maximum for higher N/Z reaction, i.e.,
$^{60}$Ca+$^{60}$Ca. Since we are having isotopes of Ca, so
Coulomb potential will be same for all the three N/Z reactions. So
the isospin effects in the collective flow will be due to the
interplay of symmetry energy and isospin-dependent nucleon-nucleon
cross section. To see the effect of symmetry energy on the
collective transverse in-plane flow, we make the strength of
symmetry energy zero. The results are displayed by dashed (red)
lines. We see that when we make the strength of symmetry energy
zero, the collective transverse in-plane flow decreases in all the
three reactions. The decrease in flow is due to the fact that
symmetry energy is repulsive in nature and hence leads to positive
in-plane flow and so when we make it's strength zero, the flow
decreases.
\par
\begin{figure}[!t] \centering
\vskip 0.5cm
\includegraphics[angle=0,width=14cm]{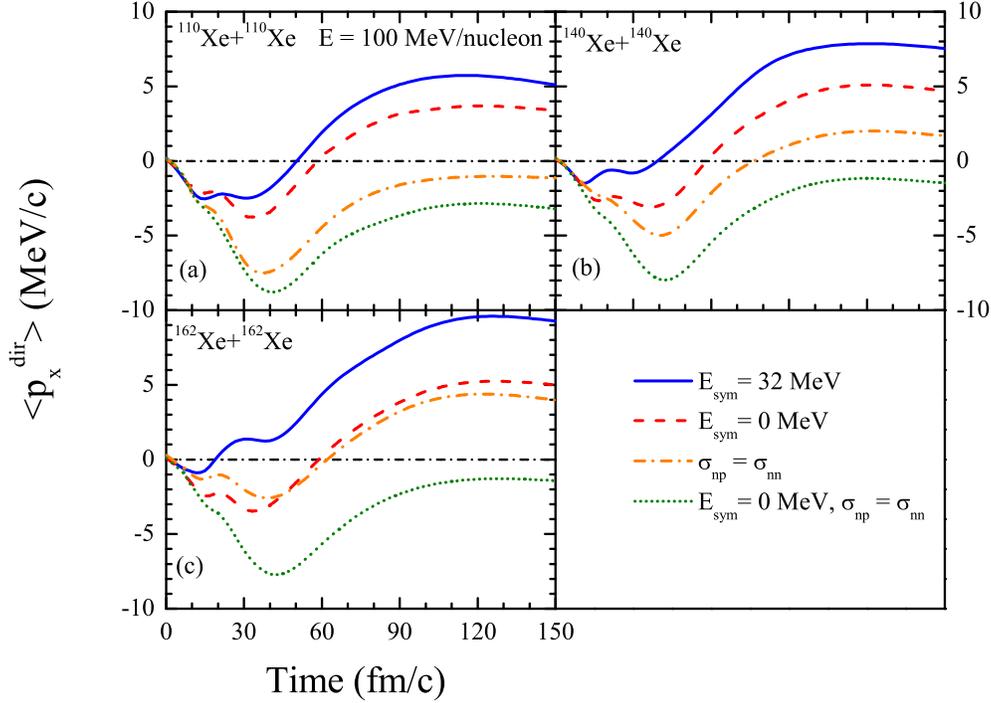}
\vskip 0.5cm \caption{(Color online) Same as Fig. 1 but for Xe+Xe
reactions.}\label{fig3}
\end{figure}
 To see the effect of isospin dependence of
nucleon-nucleon cross section, we make the cross section isospin
independent, i.e., $\sigma_{np}=\sigma_{nn}$ and calculate the
flow. The results are displayed by dash-dotted lines (orange). We
find that the flow decreases when we make the cross section
isospin independent. This is because in isospin dependent case,
the neutron-proton cross section is three times that of
neutron-neutron or proton-proton cross section. When we make the
cross section isospin independent the effective magnitude of
nucleon-nucleon cross section decreases which leads to less
transverse flow. Finally to see the combined effect of symmetry
energy and isospin dependence of cross section, we make both the
strength of symmetry energy zero and cross section to be isospin
independent, simultaneously. The results are displayed by green
lines. We find that the maximum decrease in flow is for
$^{60}\textrm{Ca}$+$^{60}\textrm{Ca}$ which goes on decreasing as
we are moving to symmetric systems.
\par

\begin{figure}[!t] \centering
 \vskip -1cm
\includegraphics[width=14cm]{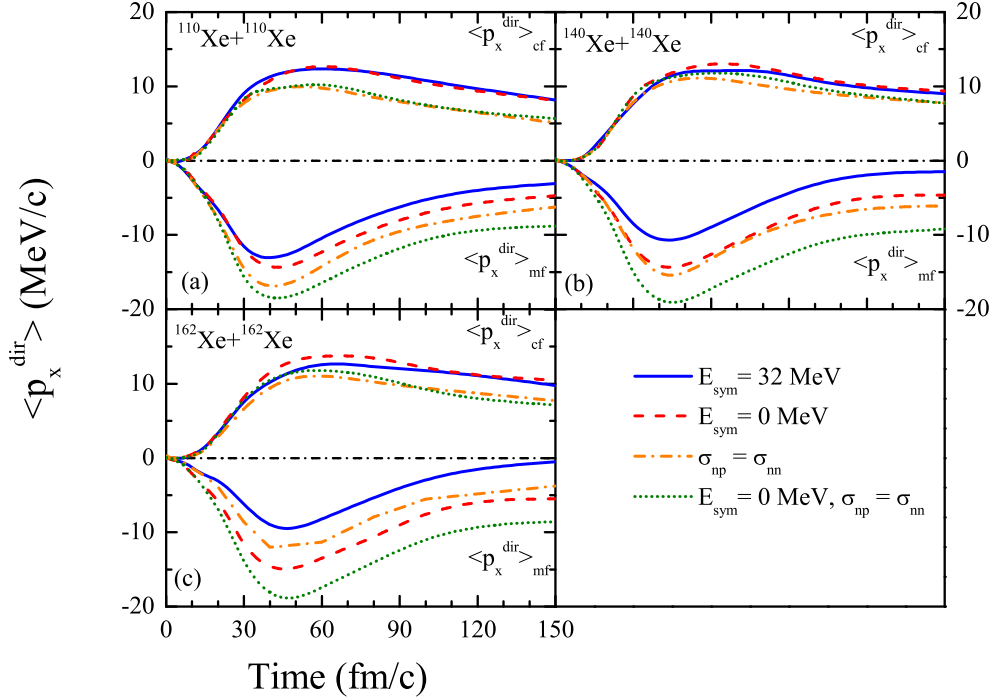}
\caption{(Color online) Same as Fig. 2 but for Xe+Xe
reactions.}\label{fig4}
\end{figure}

In fig. 2, we display the mean-field
($<p_{x}^{\textrm{dir}}>_{mf}$) and collision
($<p_{x}^{\textrm{dir}}>_{cf}$) contribution to total
$<p_{x}^{\textrm{dir}}>$ for the reactions of $^{40}$Ca+$^{40}$Ca,
$^{52}$Ca+$^{52}$Ca and $^{60}$Ca+$^{60}$Ca. We see that collision
flow remains positive throughout the reaction whereas the flow due
to mean field remains negative. We see that when we make the
strength of symmetry energy zero, the flow due to mean-field
changes and becomes less whereas the collision flow remains almost
constant. Similarly the decrease in the flow due to isospin
dependent cross section and both symmetry energy and isospin
dependent cross section is mainly reflected in mean field flow. We
find that the effect of symmetry energy is completely reflected in
$<p_{x}^{\textrm{dir}}>_{mf}$ (see red and blue lines of
$<p_{x}^{\textrm{dir}}>_{cf}$) whereas the effect of cross section
is reflected both in $<p_{x}^{\textrm{dir}}>_{mf}$ and
$<p_{x}^{\textrm{dir}}>_{cf}$ (see blue and orange lines in
$<p_{x}^{\textrm{dir}}>_{cf}$).

 \begin{figure}[!t] \centering
 \vskip -1cm
\includegraphics[width=10cm]{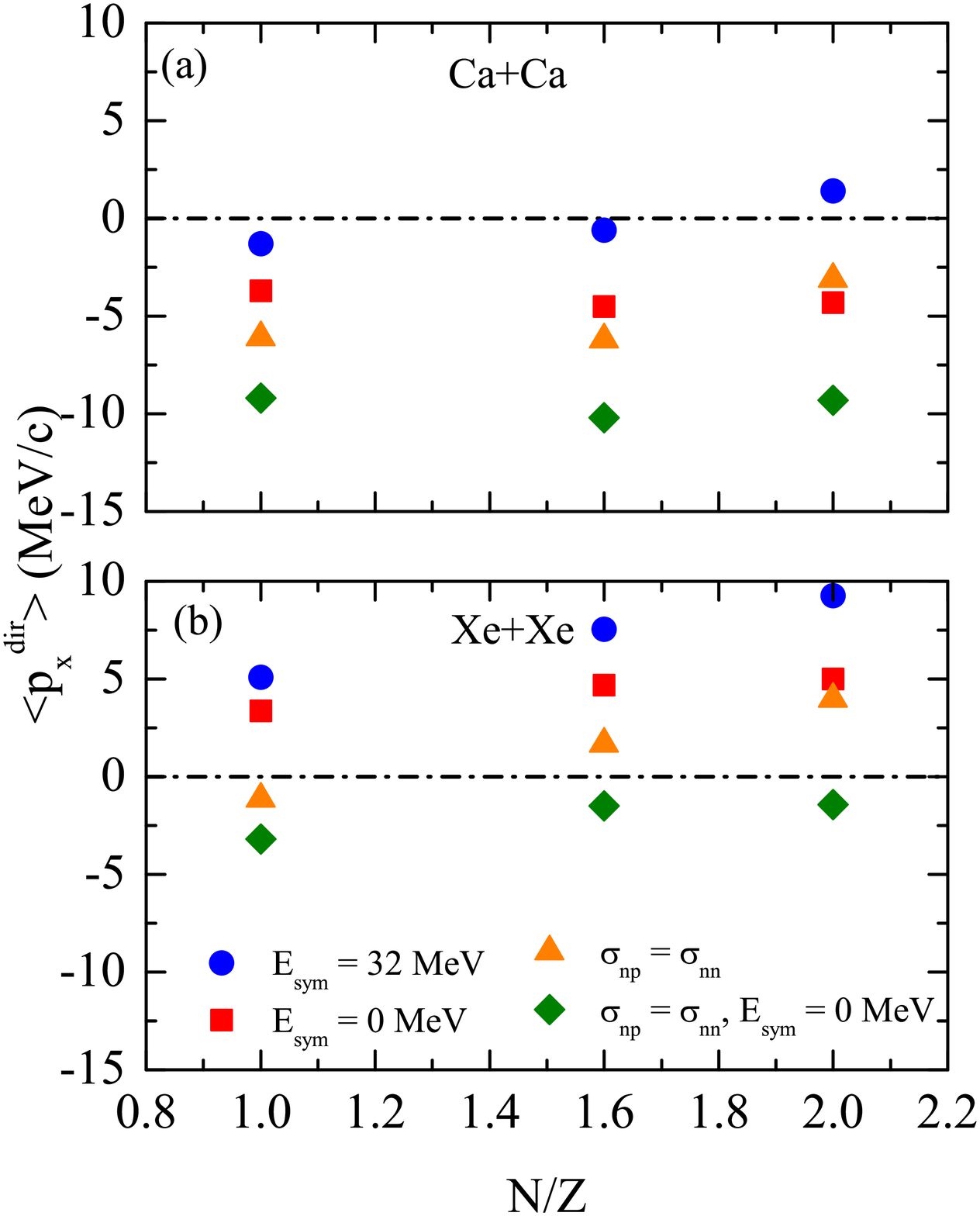}
\caption{(Color online) The N/Z dependence of transverse in-plane
flow for the reactions of Ca+Ca (upper panel) and Xe+Xe (lower).
Various symbols are explained in the text.}\label{fig5}
\end{figure}

\begin{figure}[!t] \centering
 \vskip -1cm
\includegraphics[width=10cm]{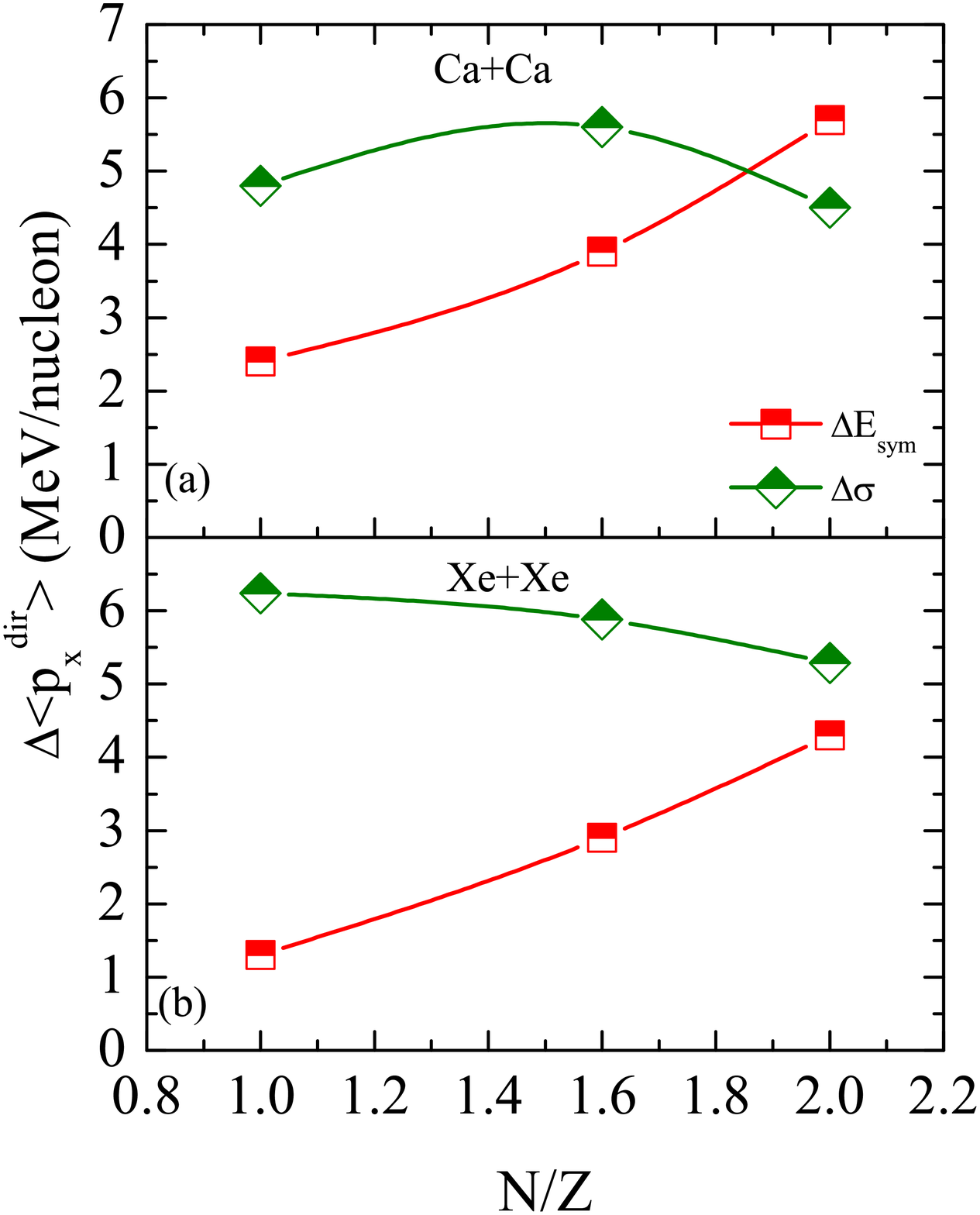}
\caption{(Color online) The difference in the
$<p_{x}^{\textrm{dir}}>$ for the reactions of Ca+Ca (upper panel)
and Xe+Xe (lower) between calculations with and without symmetry
energy and isospin-dependent and independent cross section.
Symbols are explained in the text.}\label{fig6}
\end{figure}

\par
In fig. 3, we display the time evolution of
$<p_{x}^{\textrm{dir}}>$ for the reactions of
$^{110}$Xe+$^{110}$Xe, $^{140}$Xe+$^{140}$Xe and
$^{162}$Xe+$^{162}$Xe. We see that the transverse flow is more
than that for the Ca+Ca reactions. To see the role of symmetry
energy in heavier systems like Xe+Xe, we again make the strength
of symmetry energy zero. The results are displayed by red dashed
lines. We find that the flow decreases when we make the strength
of symmetry energy zero as for the case of Ca+Ca reactions.
Similarly, the flow decreases further when me make the cross
section isospin independent. This indicates that the isospin
dependence of cross section is dominant than symmetry energy in
the isospin effects in transverse flow. The flow decreases further
when we simultaneously reduce the strength of symmetry energy zero
and make the cross section isospin independent as for the
reactions of Ca+Ca.
 \par

  In fig. 4, we divide the $<p_{x}^{\textrm{dir}}>$ into mean field
  ($<p_{x}^{\textrm{dir}}>_{mf}$) and collision contribution ($<p_{x}^{\textrm{dir}}>_{cf}$)
  for the reactions of Xe+Xe. We find that the effect of symmetry
  energy is reflected in mean field part $<p_{x}^{\textrm{dir}}>_{mf}$
whereas the effect of nucleon-nucleon cross section is reflected
in both $<p_{x}^{\textrm{dir}}>_{mf}$ and
$<p_{x}^{\textrm{dir}}>_{cf}$.
\par
In fig. 5, we display the N/Z dependence of
$<p_{x}^{\textrm{dir}}>$ for the reactions of Ca+Ca (upper panel)
and Xe+Xe (lower). Circles and squares represent the calculations
with and without symmetry energy, respectively. Triangles and
diamonds represent the calculations with isospin independent cross
section and both isospin independent cross section and without
symmetry energy, respectively. We see that for both the reactions
of Ca+Ca as well as for Xe+Xe, the transverse flow increases with
N/Z of the system (circles) due to the enhanced effect of symmetry
energy in higher N/Z systems. The decrease in the flow when we
reduce the strength of symmetry energy to zero also increases with
N/Z (see squares). We also see that decrease in flow with isospin
independent cross section (triangles) is almost the same for all
the three N/Z reactions foe both the systems. This indicates that
the role of isospin dependence of cross section is uniform
throughout the N/Z series. This is also predicted in Ref.
\cite{gaum410} where the N/Z dependence of energy of vanishing
flow is studied.

\par

In fig. 6, we display the change in flow between calculations with
and without symmetry energy (red symbols) and isospin-dependent
and isospin independent cross section (green symbols). We see that
change in the flow increases sharply with N/Z of the system for
both Ca+Ca (upper panel) and Xe+Xe (lower) reactions, whereas the
change in flow due to isospin dependence of cross section is
almost constant with N/Z. Also, the change in flow due to symmetry
energy is lore in case of Ca+Ca, indicating the greater role of
symmetry energy in lighter systems. Thus lighter systems can act
as better probes to constrain symmetry energy.
\par
In summary,we have studied the transverse flow for systems having
various
 N/Z ratios. We found the transverse flow is sensitive to N/Z
 ratio and, in fact, increases with N/Z of the system. The
 relative contribution of symmetry energy and isospin dependence
 of nucleon-nucleon cross section is also investigated. We also
 found the greater sensitivity of symmetry energy in lighter
 systems as compared to the heavier ones.
 \par
This work has been supported by a grant from Centre of Scientific
and Industrial Research (CSIR), Govt. of India and Indo-French
Centre For The Promotion Of Advanced Research (IFCPAR) under
project no. 4104-1.

\end{document}